\documentclass[prc,twocolumn,epsfig,nofootinbib,floatfix]{revtex4}
\usepackage{graphics}
\usepackage{epsfig}
\usepackage{amsfonts}
\usepackage{amsmath}
\usepackage{bm}% bold math

\begin{document}
\title{
SKYRME-RPA DESCRIPTION OF DIPOLE GIANT RESONANCE IN HEAVY AND SUPERHEAVY
NUCLEI}
\author{W. Kleinig$^{1,2}$, V.O. Nesterenko$^1$, J. Kvasil$^3$,
P.-G. Reinhard$^4$ and P. Vesely$^3$}
\affiliation{$^1$ Laboratory of Theoretical Physics,
Joint Institute for Nuclear Research, Dubna, Moscow region, 141980, Russia}
\email{nester@theor.jinr.ru}
\affiliation{$^2$ Technische Universit\"at Dresden, Inst. f\"ur Analysis,
   D-01062, Dresden, Germany}
%\author{J. Kvasil, P. Vesely}
\affiliation{$^3$ Institute of Particle and Nuclear Physics, Charles
University, CZ-18000, Praha 8, Czech Republic}
%\author{P.-G. Reinhard}
\affiliation{$^4$ Institut f\"ur Theoretische Physik II,
Universit\"at Erlangen, D-91058, Erlangen, Germany}

\date{\today}

\begin{abstract}
The E1(T=1) isovector dipole giant resonance (GDR) in heavy and super-heavy
deformed nuclei is analyzed over a sample of 18 rare-earth nuclei, 4 actinides
and three chains of super-heavy elements (Z=102, 114 and 120). Basis of the
description is self-consistent separable RPA (SRPA) using the Skyrme force
SLy6.  The self-consistent model well reproduces the experimental data
(energies and widths) in the rare-earth and actinide region.  The trend of the
resonance peak energies follows the estimates from collective models, showing a
bias to the volume mode for the rare-earths isotopes and a mix of volume and
surface modes for actinides and super-heavy elements.  The widths of the GDR
are mainly determined by the Landau fragmentation which in turn  is found
to be strongly influenced by deformation. A deformation
splitting of the GDR can contribute about one third to the width and about 1
MeV further broadening can be associated to mechanism beyond the mean-field
description (escape, coupling with complex configurations).
\end{abstract}


\maketitle

\section{Introduction}
\label{sec:introduction}

The isovector giant dipole resonance (GDR) is a most prominent and much studied
excitation mode of nuclei, see e.g. \cite{BF_75,W_87}. Nonetheless, it remains
a subject of actual interest as there are many aspects which deserve more
detailed investigations as, e.g., photo-excitation cross sections in exotic
nuclei which play a role in astrophysical reaction chains \cite{astrochains}
or isotopic trends of the GDR including the regimes of deformed nuclei. The
present paper aims at a theoretical survey of the GDR in isotopic chains of
heavy and super-heavy nuclei.

The high importance of the GDR has triggered since long many
theoretical surveys analyzing the intriguing aspects of nuclear
collective motion, starting from a purely collective description
\cite{GT_48,SJ_50} and slowly establishing a link to microscopic
models in the framework of the Random-Phase Approximation (RPA)
\cite{Row70aB,Bro71aB}. The theoretical description has much developed
over the years. The majority of RPA investigations in the past
employed shell model potentials plus an effective residual interaction
(Migdal theory) \cite{Goe82aR,Spe91aR,Ber94aB}. In the meantime,
self-consistent nuclear models have been steadily improving towards a
reliable description of nuclear structure and excitations, for reviews
see, e.g., \cite{Ben,Vre05aR,Sto07aR}.  These models belong to the
class of density functional methods which aim at a universal energy
functional for a given system from which all static and dynamics
equations are derived in a strictly variational frame
\cite{Dre90aB0,Pet91aB}. Such density functional models, being rather
universal by construction, are promising for investigation in exotic
areas, e.g. for r-process, drip-line and super-heavy nuclei.
The studies in this paper are based on the Skyrme functional which has been
introduced in \cite{Skyrme,Vau} and extended to a dynamical description shortly
after \cite{Engel_75,Shl75a,Kre77a}.  The performance of self-consistent RPA
calculations using the Skyrme functional had been tested systematically in
\cite{Rei92b,Rei99a} and it was found that one can have a reliable description
when taking care to chose an appropriate parametrization.

Systematic scans through the isotopic landscape and the study of
exotic nuclei run over many deformed nuclei.  Fully fledged RPA
calculations for deformed nuclei are feasible \cite{Mar05b}, but
extremely time consuming, not suited for systematic investigations.
Super-heavy nuclei are especially demanding due to a coexistence of
two obstacles, large size and deformation.  Accurate but less
demanding RPA techniques are needed.  To that end, the separable RPA
(SRPA) based on the Skyrme functional was recently developed
\cite{nest_02,nest_06}.  The self-consistent factorization of the
residual interaction in SRPA reduces the computational expense
dramatically and so gives way to systematic explorations of nuclear
giant resonances in both spherical and deformed nuclei
\cite{nest_02}-\cite{nest_kazim}. In this paper we concentrate on
the isovector (T=1) electric GDR.

As was shown in our previous studies \cite{nest_02}-\cite{nest_kazim},
SRPA provides an accurate description of the GDR in spherical and
deformed nuclei. We obtained good agreement with experiment for
$^{154}$Sm, $^{238}$U and Nd isotopes with A=142, 144, 146, 148, 150.
Eight different Skyrme forces were checked in these investigations.
In this paper, we aim at a large systematics over the isotopic
landscape and decide for one parameterization, namely SLy6 \cite{Sly6}
which was found to provide a satisfying description of the GDR for
spherical and deformed nuclei.
In a first step, SRPA results for the GDR will be compared with all
available experiment data in rare-earth and actinide regions. In
particular, we consider nuclei $^{156,160}$Gd, $^{166,168}$Er, $^{170,
172, 174, 176}$Yb, $^{176, 178, 180}$Hf, $^{182, 184, 186}$W, $^{186,
188, 190, 192}$Os, $^{232}$Th and $^{234, 236, 238}$U.  Basic
characteristics (energy centroid, width, deformation splitting) and
their trends with system size will be analyzed.

In a second step, we will investigate the GDR in super-heavy nuclei for the
three isotopic chains: Nobelium with Z=102 (A=242, 248, 254, 262, 270), Z=114
(A=264, 274, 284, 294, 304) and Z=120 (A=280, 288, 294, 304, 312).  As
discussed below, this set covers most of the important mass regions and so is
sufficiently representative. The main features of the GDR are analyzed and
compared with those in rare-earth and actinide region.

The paper is organized as follows.  The calculation scheme, methods
of analysis and choice of the isotopes are sketched in Sec.
\ref{sec:calc_scheme}. In Sec. \ref{sec:results} the results of the
calculations for the GDR in rare-earth, actinide and super-heavy
nuclei are discussed.  A summary is done in Sec. \ref{sec:summary}.

\section{Framework}
\label{sec:calc_scheme}

The calculations are performed in the framework of SRPA
\cite{nest_kazim} with the Skyrme force SLy6 \cite{Sly6} which
provides a satisfying description of the GDR for heavy nuclei
\cite{nest_06,nest_07,Rila,nest_kazim}. The contribution to the
residual interaction from the time-odd current density, Coulomb and
pairing (at the BCS level) are taken into account \cite{nest_06}. The
calculations employ a cylindrical coordinate-space grid with the mesh
size 0.7 fm. The calculation box has 24-35 mesh points depending on
the nuclear size and deformation. For generators of the separable
interaction, we use four input operators, $rY_{1\mu}$ and  $j(q_ir)Y_{1\mu}$,
$i=1,2,3$, with $q_i$ chosen following the prescription
\cite{nest_02}. These operators are shown to give
accurate results in spherical nuclei \cite{nest_02}. Besides, the operator
$r^3 Y_{3\mu}$ is added to take into account the multipole mixing of
excitations with the same projection $\mu$ and space parity $\pi$.

For all nuclei, the equilibrium quadrupole deformations are found by
minimization of the total energy. The deformations are characterized
by the charge quadrupole moments $Q_{2}$ and related dimensionless
parameters $\beta_2$ as
\begin{equation}\label{eq:quad_def}
  Q_{2} = \int d\vec r \rho_p(\vec r) r^2 Y_{20},
  \quad
  \beta_2 = \sqrt{\frac{\pi}{5}} \frac{Q_2}{Z <r^2>_p}
\end{equation}
where $\rho_p(\vec r)$ is the proton density in the ground state,
$<r^2>_p=\int d{\vec r}\rho_p(\vec r) r^2/Z$ is the r.m.s. proton radius and
$Z$ is the number of protons.

The calculations use a large basis space of single-particle states to
expand the SRPA operators with two-quasiparticle states extending up
to $\sim 65$ MeV, see Table I.  The energy-weighted sum rule
\begin{equation}\label{ewsr_1}
EWSR(T=1, \lambda=1)=9\frac{(\hbar e)^2}{8\pi m^*_1}  \frac{NZ}{A} \; .
\end{equation}
is then  exhausted by 92-95$\%$. This sum rule includes the isovector
effective mass $m^*_1$ as the velocity-dependent terms are involved to
the calculations, see discussion in \cite{nest_kazim}. In SLy6 we have
actually $m^*_1/m$=0.80.

Since SRPA includes the dipole time-odd momentum-like operators
\cite{nest_06}, the center-of-mass mode should, in principle, be placed
correctly at zero energy. However, the finite computational box breaks
translational invariance such that the spurious isoscalar E1 strength
becomes concentrated at the region of 2-3 MeV.  That is safely below
the GDR which we are studying here.

The energy-weighted isovector dipole strength function is computed as
\begin{eqnarray}
  S(E 1\mu ; \omega)
  &=&
  \sum_{\nu}
  \omega_{\nu}\langle\Psi_\nu|\hat{E}_{1\mu}|\Psi_0\rangle
  \zeta(\omega - \omega_{\nu})
  \quad,
\label{eq:strength_function}
\\
  \hat{E}_{1\mu}
  &=&
  \frac{N}{A}\sum_{p=1}^Z r_p Y_{1\mu}(\Omega_p)
  -
  \frac{Z}{A}\sum_{n=1}^N r_n Y_{1\mu}(\Omega_n)
  \quad,
\nonumber
\end{eqnarray}
smoothed by the Lorentz function
\begin{equation}
  \zeta(\omega - \omega_{\nu}) =
  \frac{1}{2\pi}\frac{\Delta}{(\omega- \omega_{\nu})^2+\frac{\Delta^2}{4}}
\end{equation}
with the averaging parameter
$\Delta$=2 MeV in most of the calculations.  That averaging width was
found to be optimal for the comparison with experiment and simulation
of broadening effects beyond SRPA (escape widths, coupling with
complex configurations).  Further, $\Psi_0$ is the ground state, $\nu$
runs over the RPA spectrum with eigenfrequency $\omega_{\nu}$ and
eigenstate $|\Psi_\nu>$.

\begin{table}
\caption{\label{tab:config_space}
Characteristics of the configuration space used in the calculations:
minimal $E_{min}$ and maximal $E_{max}$ single-particle
energies, number of the single-particle levels $K$ for protons and
neutrons, number of two-quasiparticle dipole states $N_{2qp}$ (for branches
$\mu=0$ and 1 altogether). See text for more details.
}
\begin{tabular}{|c|c|c|c|c|c|c}
\hline
Nucleus & \multicolumn{2}{|c|}{$E_{min}, \quad E_{max}$}
& \multicolumn{2}{|c|}{K}  & $N_{2qp}$ \\
\cline{2-5}
& Z & N & Z & N & \\
\hline
 $^{154}$Gd & -45.4, +20.2  & -57.2, +17.0 & 252 & 308 & 4720 \\
 $^{238}$U  & -42.7, +19.3  & -58.0, +14.8 & 307 & 393 & 6860  \\
 $^{294}$120 & -36.4, +20.9 & -58.7, +14.1 & 360  & 426 & 8720 \\
\hline
\end{tabular}
\end{table}

To estimate the resonance energy centroid $E$, width $\Gamma$, and
deformation splitting $\Delta E$, the following prescriptions are
applied. To determine $E$, the energy interval around the resonance
with the strength larger than 10$\%$ of the maximal value is used and
the centroid of the strength inside this interval is determined. The
same method is implemented to find centroids $E_0$ and $E_1$ of
$\mu=0$ and 1 branches separately. Then the deformation splitting
$\Delta E=E_1-E_0$ is obtained. The width $\Gamma$ is determined at a
half-maximum of the resonance.

The experimental data for the GDR
\cite{atlas,Gur_NPA_81,Varl_BRAS_03,Gor_YF_76,Gor_VTYF_76,Gor_YF_77,
Ber_PRC_79,Gur_NPA_76,Ber_PRC_86,Cald_PRC_80}
include photoabsorption
\begin{equation}
\sigma = (\gamma,n)+(\gamma,p)+(\gamma,np)+(\gamma,2n)+(\gamma,d)+...+(\gamma,f) \; ,
\nonumber
\end{equation}
neutron yield
\begin{equation}
\sigma = (\gamma,n)+(\gamma,np)+2(\gamma,2n)+3(\gamma,3n)+...+(\gamma,f) \; ,
\nonumber
\end{equation}
and neutron product
\begin{equation}
\sigma = (\gamma,n)+(\gamma,np)+(\gamma,2n)+(\gamma,3n)+...+(\gamma,f) \; .
\nonumber
\end{equation}
The photoabsorption data are preferable as involving all
the main decay channels and so most corresponding to the strength function
(\ref{eq:strength_function}).
If the photoabsorption data are absent, the neutron yield and neutron product
can be also used for a rough comparison since these data include most of the
main channels. However, one should take into account that
both neutron yield and product omit $(\gamma,p)$ channel and so can underestimate
the strength and change the resonance gross-structure.
Besides,  the neutron yield amplifies the neutron contributions
$(\gamma,2n), (\gamma,3n), ...$ and hence the right wing of the resonance.
%
% Figure 1
%%%%%%%%%%%%
\begin{figure}[t]
\includegraphics[height=6.5cm,width=7cm]{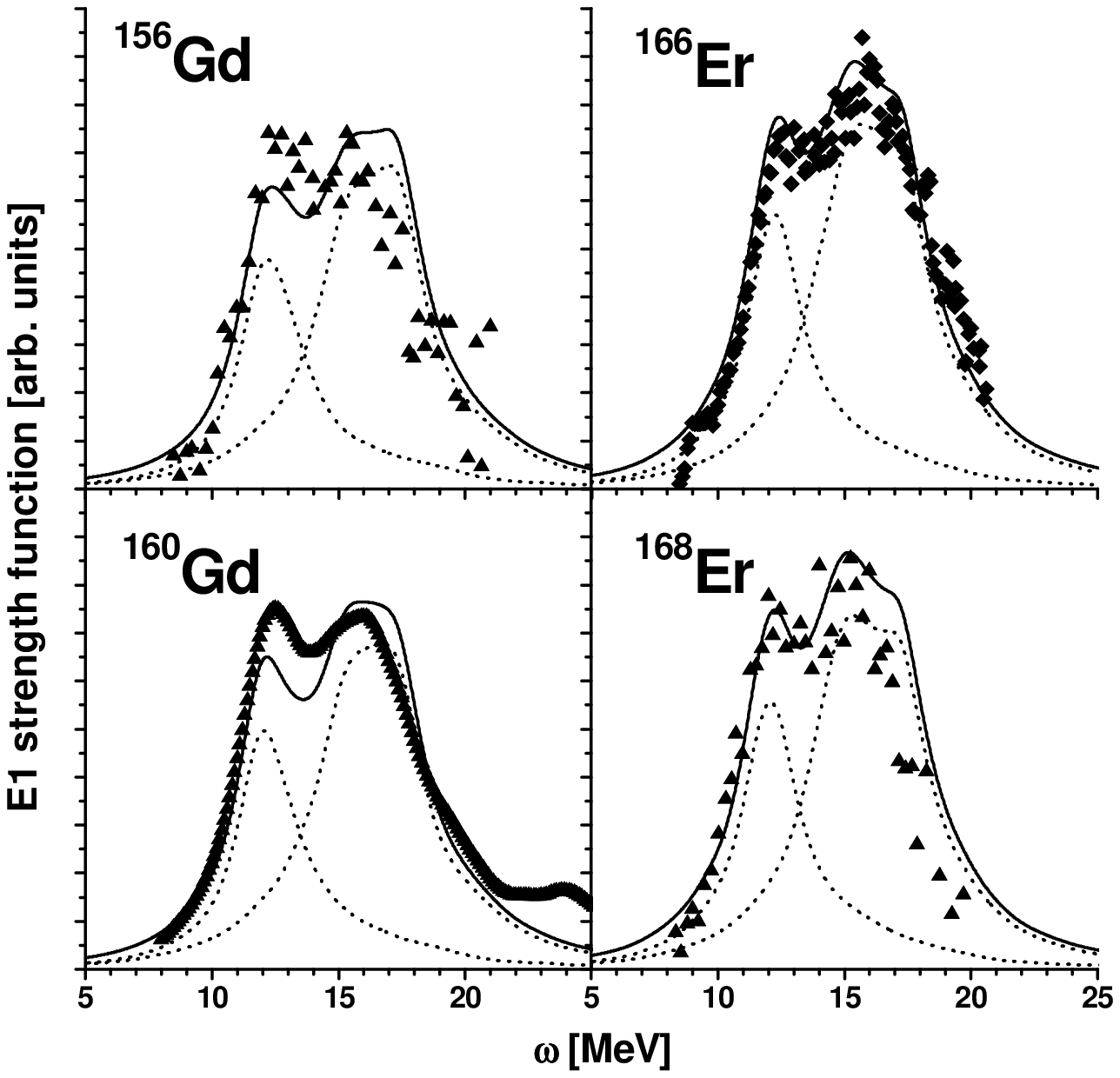}
%\vspace{0.4cm}
\caption{\label{fig:gd_er}
Isovector dipole strength
in $^{156,160}$Gd and $^{166,168}$Er.
The calculated strength (solid curve) is compared with
the experimental data for total photoabsorption
\protect\cite{Gur_NPA_81,Varl_BRAS_03}(triangles)
and neutron product \protect\cite{Gor_YF_76}(rhombus).
The branches of the resonance with $\mu = 0$ (left small)
and $\mu = 1$ (right large) are represented by dotted curves.
The deformations are $\beta_2$=0.347, 0.359 and 0.348, 0.353,
respectively.
}
\end{figure}
In what follows, we use experimental data  for
photoabsorption in
$^{156}$Gd \cite{Gur_NPA_81},
$^{160}$Gd \cite{Varl_BRAS_03},
$^{168}$Er \cite{Gur_NPA_81},
$^{174}$Yb  \cite{Gur_NPA_81},
$^{178,180}$Hf \cite{Gur_NPA_81},
$^{182,184,186}$W \cite{Gur_NPA_81},
$^{232}$Th \cite{Gur_NPA_76},
$^{238}$U \cite{Gur_NPA_76},
neutron yield in
$^{170,172,176}$Yb \cite{Gor_VTYF_76},
$^{186,188,190,192}$Os \cite{Ber_PRC_79},
neutron products $(\gamma,n)+(\gamma,np)+(\gamma,2n)$ in
$^{166}$Er \cite{Gor_YF_76},
$^{176}$Hf \cite{Gor_YF_77}
and $(\gamma,n)+(\gamma,np)+(\gamma,2n)+(\gamma,f)$ in
$^{234}$U \cite{Ber_PRC_86}, $^{236}$U  \cite{Cald_PRC_80}.

In rare-earth and actinide regions, we consider all nuclei for which
reasonable GDR experimental data exist (for exception of Nd and Sm
isotopes already explored in our previous papers
\cite{nest_06,nest_07,Rila,nest_kazim}).  In super-heavy nuclei, we look
at three isotopic chains: Nobelium $Z$=102 (A=242, 248, 254, 262, 270),
$Z$=114 (A=264, 274, 284, 294, 304) and $Z$=120 (A=280, 288, 294, 304, 312).
As can be seen from \cite{baran}, these chains cover most interesting
mass and deformation regions. Indeed, they involve the onset
($Z$=102), the center ($Z$=114) and the upper end ($Z$=120) of the
super-heavy region.  Every chain extends through the whole neutron
interval at a given $Z$.  Different deformation regions are covered.
For $Z$=102 we deal with well deformed nuclei and small variation of
the quadrupole deformation. The chains $Z$=114 and 120 show strong
variations of the deformation with a strong decrease with increasing
$N$, i.e. when moving towards the magic neutron number $N$=184
\cite{Ben99a}.
The proton numbers $Z$=114 and 120 are tentatively magic
\cite{Ben01a,Ben} such that neutron shell structure acquires a
decisive weight for sphericity or deformation.

\section{Results and discussion}
\label{sec:results}
\subsection{Rare-earth and actinide nuclei}

% Figure 2
%%%%%%%%%%%%
\begin{figure}
\includegraphics[height=6.5cm,width=7cm]{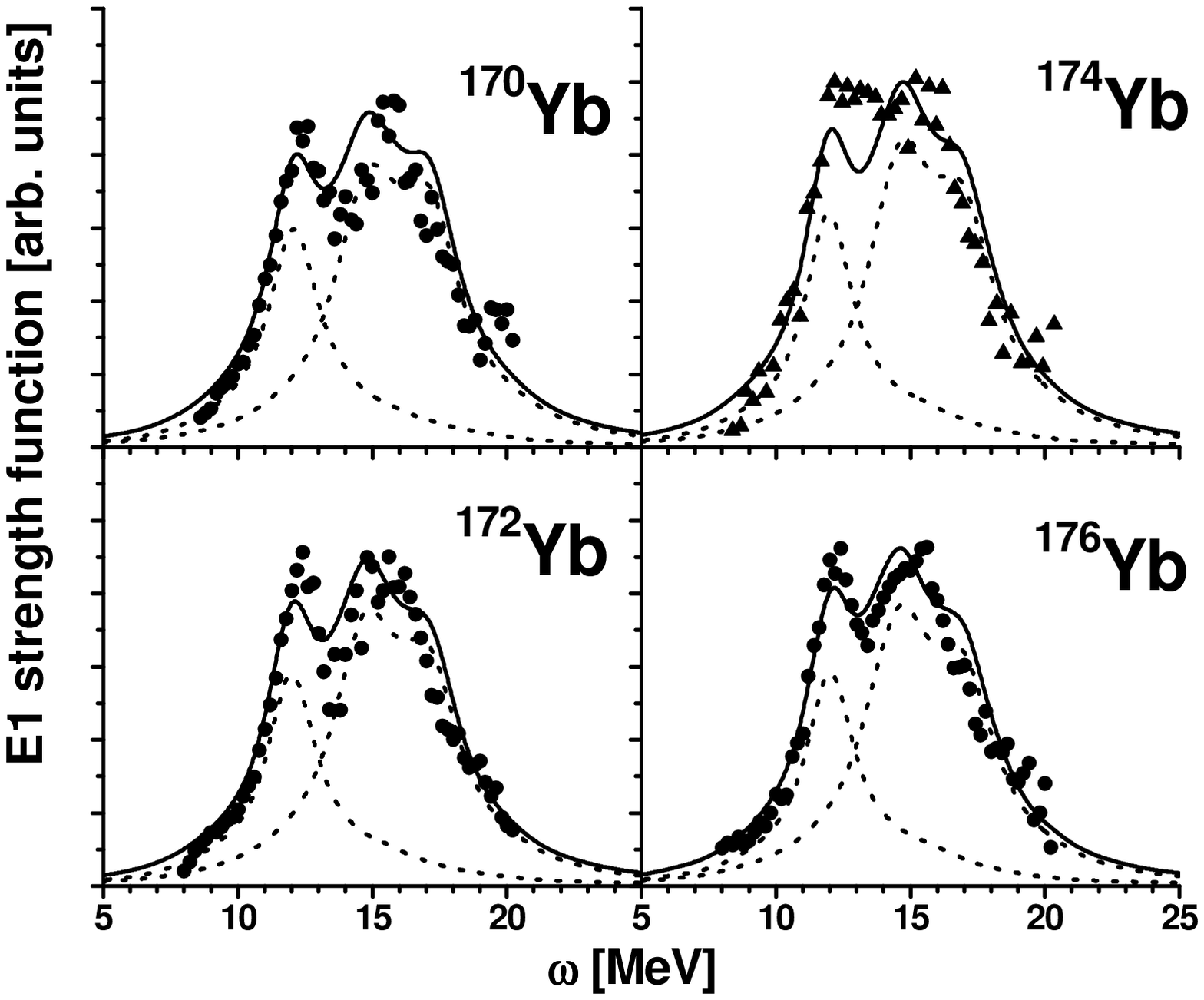}
\vspace{0.3cm}
\caption{\label{fig:yb}
Same as in Fig. 1 for $^{170,172,174,176}$Yb. The experimental data
are for neutron yield \protect\cite{Gor_VTYF_76} (closed circles)
and total photoabsorption \protect\cite{Gur_NPA_81} (triangles).
The deformation parameters are $\beta_2$=0.350, 0.347, 0.340, 0.327,
respectively.
}
\end{figure}

\begin{figure}
\includegraphics[height=7.0cm,width=7cm]{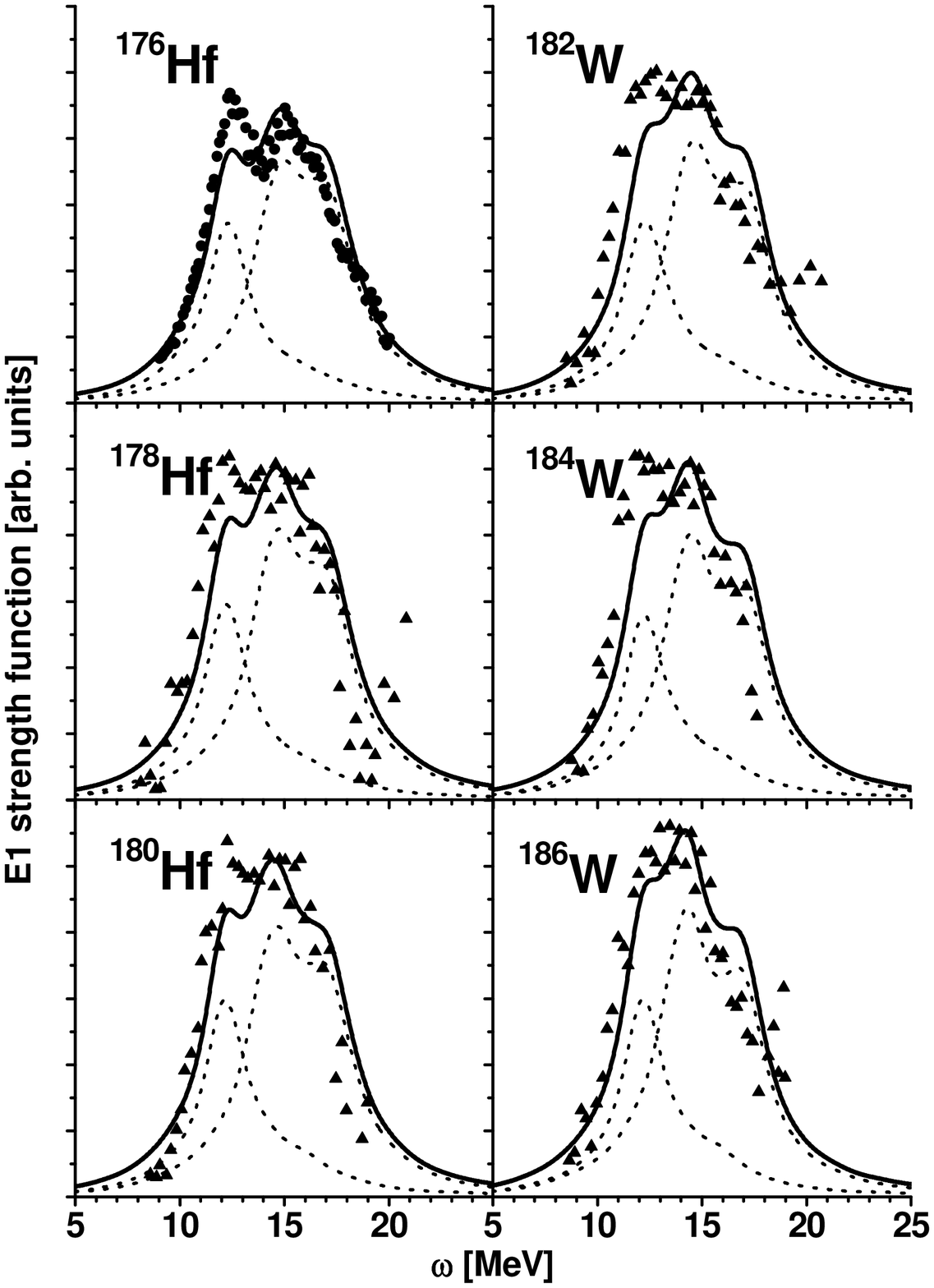}
\vspace{0.3cm}
\caption{\label{fig:hf_w}
Same as in Fig. 1 for $^{176,178,180}$Hf and $^{182,184,186}$W.
The experimental data  are for total photoabsorption
\protect\cite{Gur_NPA_81} (triangles) and neutron product
\protect\cite{Gor_YF_77} (closed circles).
The deformation parameters for $^{176,178,180}$Hf and
$^{182,184,186}$W are $\beta_2$=0.330, 0.296, 0.287
and 0.260, 0.252, 0.247, respectively.
}
\end{figure}

\begin{figure}
\includegraphics[height=6.5cm,width=7cm]{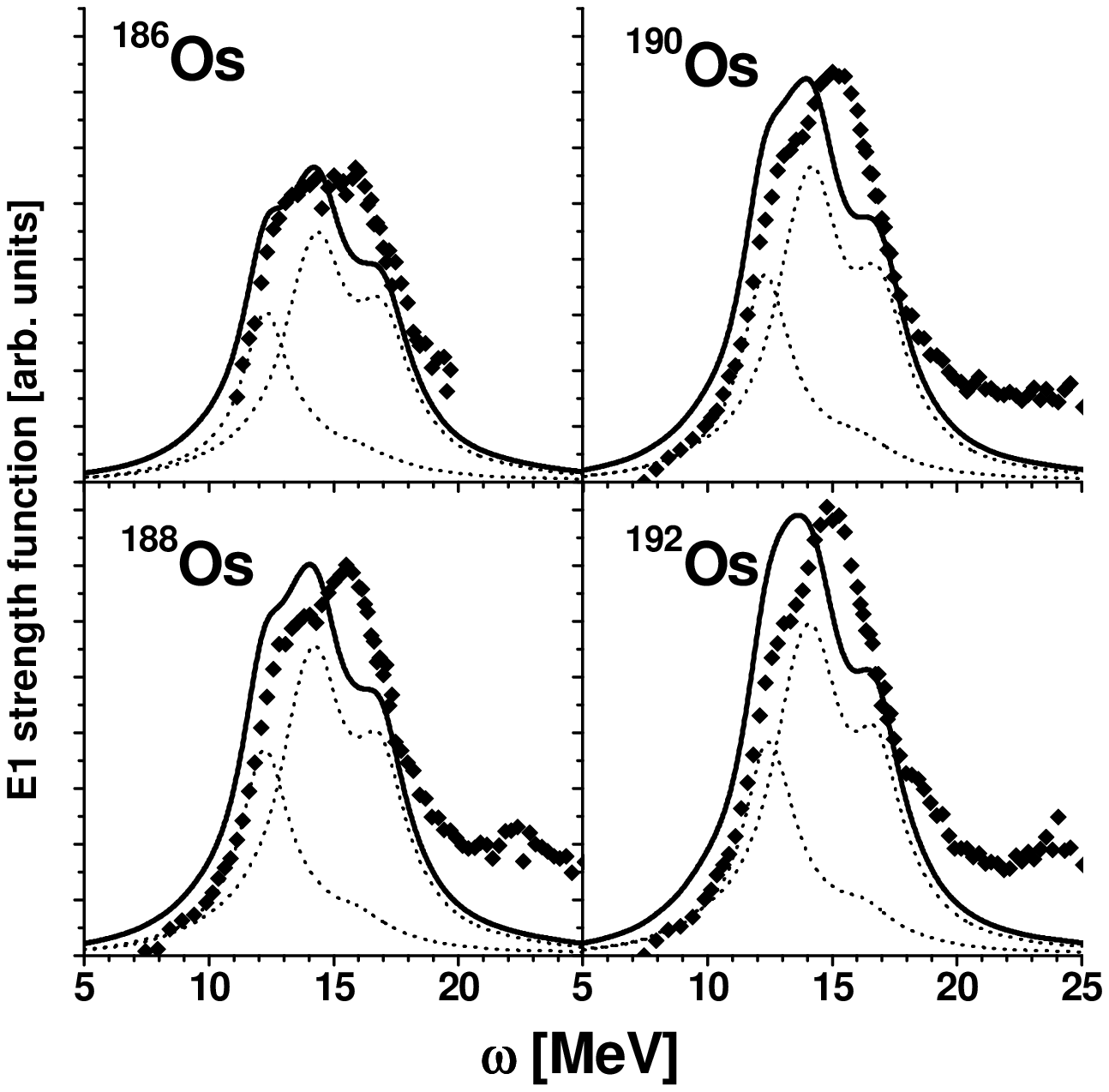}
\caption{\label{fig:os}
Same as in Fig. 1 for $^{186,188,190,192}$Os.
The experimental data  are for
neutron yield  \protect\cite{Ber_PRC_79} (rhombus).
The deformation parameters are
$\beta_2$=0.222, 0.217, 0.195, 0.172,
respectively.
}
\end{figure}

\begin{figure}
\vspace{0.5cm}
\includegraphics[height=6.5cm,width=7cm]{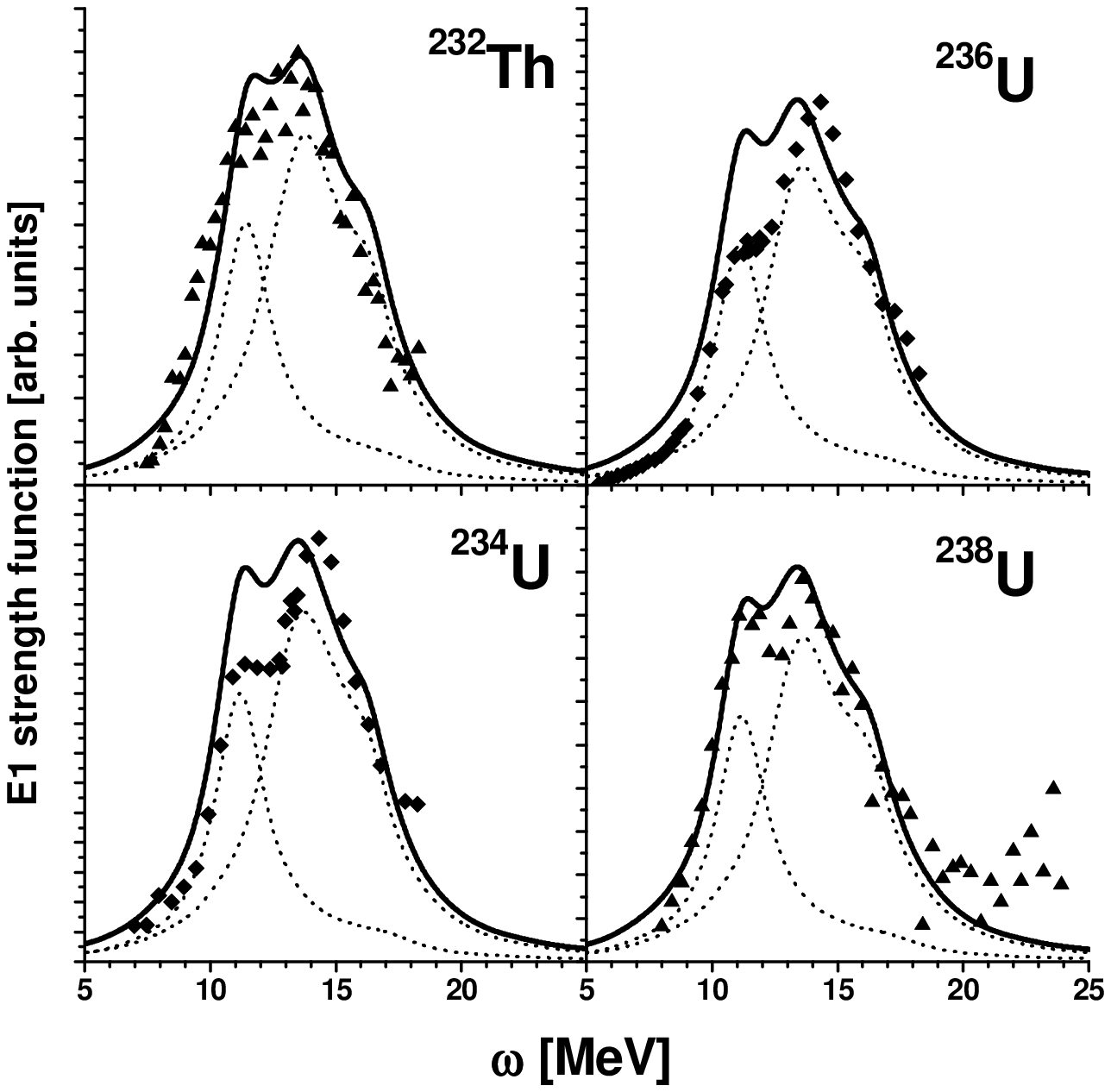}
\caption{\label{fig:th_u}
Same as in Fig. 1 for $^{232}$Th and $^{234,236,238}$U.
The experimental data  are for total photoabsorption
\protect\cite{Gur_NPA_76} (triangles) and neutron product
\protect\cite{Ber_PRC_86,Cald_PRC_80} (rhombus).
The deformation parameters for $^{232}$Th and
$^{234,236,238}$U are $\beta_2$=0.256
and 0.279, 0.286, 0.287, respectively.
}
\end{figure}

\begin{figure}
\includegraphics[height=8cm,width=8cm,angle=-90]{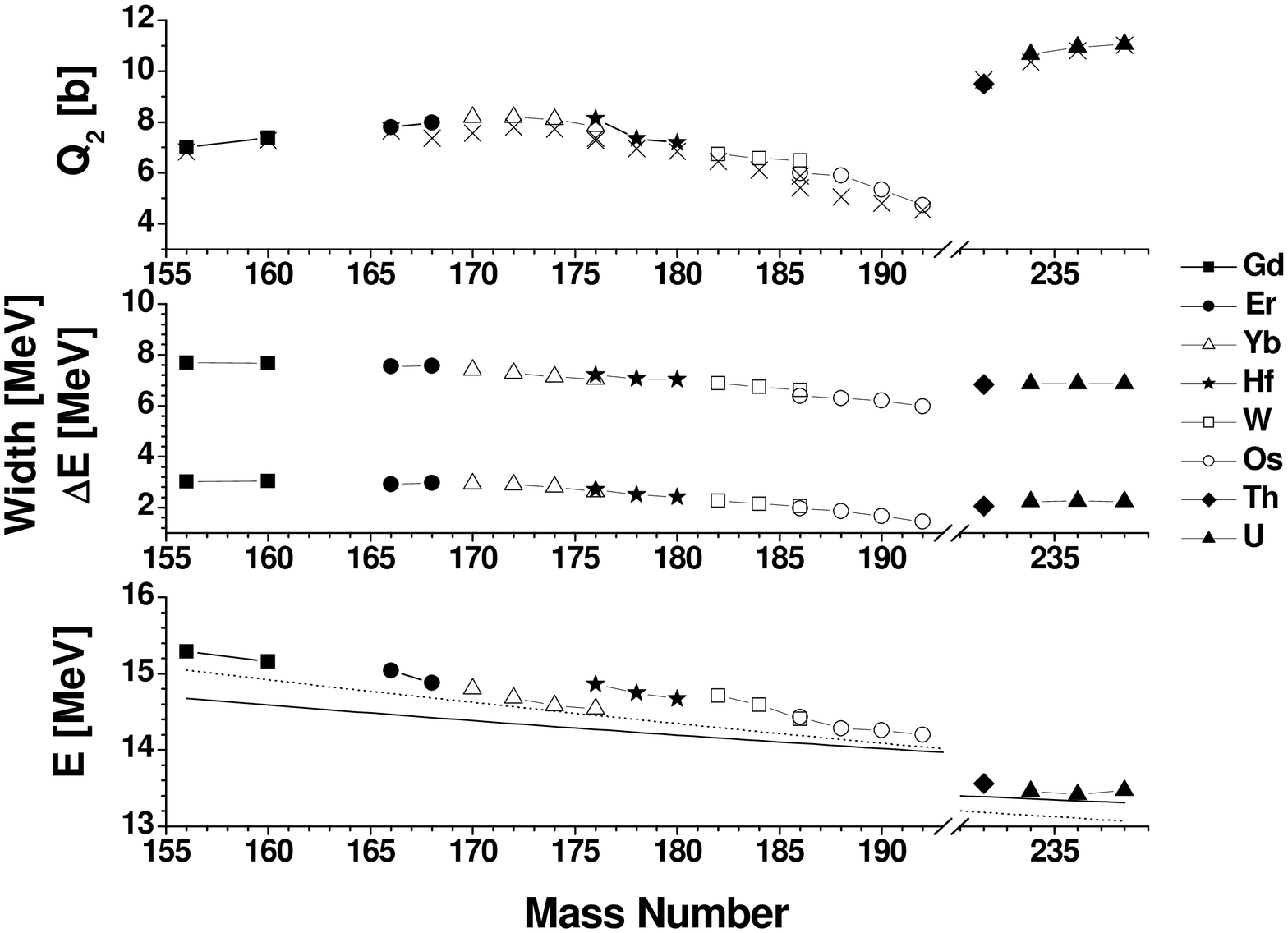}
\vspace{1cm}
\caption{\label{fig:rare_trend}
Calculated characteristics of rare-earth and actinide nuclei
as function of mass number. Upper panel: Quadrupole moments
$Q_2$ compared with experimental values (crosses)
\protect\cite{Ram_exp_Q2}. Middle panel: Widths $\Gamma$
(upper set) and deformation splittings $\Delta E$ (lower set).
Lower panel: Energy centroids as compared with the estimates
\protect(\ref{eq:SJ}) (dotted curve) and \protect(\ref{eq:BF})
(solid curve).
}
\end{figure}

Results of SRPA calculations for rare-earth and actinide nuclei are
presented in Figs. \ref{fig:gd_er}-\ref{fig:rare_trend}.  Note that
for reasons of comparison the calculated strength function is rescaled
so as to correspond roughly to the maximal magnitude of the
experimental cross section. Moreover, because of the insufficient
accuracy of the model and experimental resolution (see discussion
in the previous section) we skip here the analysis of the fine
structure which manifests itself mainly at the top of the
resonance. The main attention will be paid the resonance energy
centroids and widths.

% Figure 3
%%%%%%%%%%%%
% Figure 4
%%%%%%%%%%%%
% Figure 5
%%%%%%%%%%%%
Figures \ref{fig:gd_er} and \ref{fig:yb} show an excellent agreement
with experiment for Gd, Er and Yb isotopes.  The agreement is less
perfect for Hf, W and Os shown in Figures \ref{fig:hf_w} and
\ref{fig:os}: the calculated strength exhibits a slight ($\sim$ 0.5
MeV) down-shift in Hf and W and a larger ($\sim$ 1 MeV) up-shift in
Os. It is known that Os isotopes are soft to oblate quadrupole
deformation, which is confirmed by our calculations. However, as we
checked, SRPA calculations on top of the oblate isomer do not improve
agreement with the experiment.  The discrepancies for Os may be partly
caused by using the neutron yield experimental data. As was discussed
above, the neutron yield alone can amplify the right GDR flank, thus
resulting in some apparent up-shift as compared with the total
photoabsorption cross section.  Results for the actinides in
Fig. \ref{fig:th_u} also look encouraging.  Modest deviations in
gross-structure of $^{234,236}$U can be explained by using the neutron
product experimental data.

It is worth noting that the comprehensive analysis of experimental
data reveals noticeable (sometimes significant) deviations in GDR
measurements of different experimental groups \cite{Varl_04}.  Taking
into account these uncertainties in the data, one may consider for all
cases above the agreement with experimental data as very satisfying,
at least for energy centroids and widths.

% Figure 6
%%%%%%%%%%%%

The correlation of the quadrupole moments (\ref{eq:quad_def}) with
some resonance characteristics (width $\Gamma$, deformation splitting
$\Delta E$, energy centroid $E$) and trends of these characteristics
with the mass number $A$ are shown in Fig. \ref{fig:rare_trend}.  All
nuclei in the sample have a significant quadrupole deformation. The
calculated quadrupole moments are in excellent agreement with the
experiment \cite{Ram_exp_Q2}. In the rare-earth nuclei the moments
have a maximum in the middle of the region. Note that the
dimensionless deformation parameters $\beta_2$ as indicated in the
captions of the previous figures are maximal at the onset of the
region. The difference between $Q_2$ and $\beta_2$ maxima is related
with the nuclear mass factor in (\ref{eq:quad_def}).
The direct contribution of the deformation
splitting $\Delta E$ to the resonance width is maximal ($\sim 40\%$)
in the first half of the region ($A < 176$) and then slowly decreases
to $37-34\%$ for Hf, $34-31\%$ for W and $31-24\%$ for
Os. Furthermore, $\Delta E$ is a bit increased in actinides where it
reaches $30-33\%$. This trend obviously correlates with $\beta_2$.
See also the detailed discussion of the GDR width and structure in
section \ref{sec:width}.

In Figure \ref{fig:rare_trend}c, the resonance energy is compared with
empirical estimates based on Steinwedel-Jensen (SJ) \cite{SJ_50}
\begin{equation}\label{eq:SJ}
E_{SJ}=81 A^{-1/3} \text{MeV}
\end{equation}
and Berman-Fultz (BF) \cite{BF_75,W_87}
\begin{equation}\label{eq:BF}
  E_{BF}=(31.2A^{-1/3}+20.6A^{-1/6}) \text{MeV}
\end{equation}
collective models.
BF takes into account both volume and surface contributions and treats
the dipole resonance as a combination of Steinwedel-Jensen
\cite{SJ_50} and Goldhaber-Teller (GT) \cite{GT_48} scenarios.  The
calculated energies are closer to the SJ estimate (\ref{eq:SJ}) in the
rare-earth region and to the BF estimate (\ref{eq:BF}) in
actinides. As shown below, the BF estimate  is also more
appropriate in super-heavy nuclei. This agrees with a commonly
accepted view that in heavy nuclei neither the density gradient ($\sim
A^{-1/3}$) nor the nuclear surface impact ($\sim A^{-1/6}$) dominate
the restoring force \cite{W_87,MS_77,Bra83b}.

\subsection{Super-heavy nuclei}

\begin{figure}
\includegraphics [height=12.0cm,width=9cm]{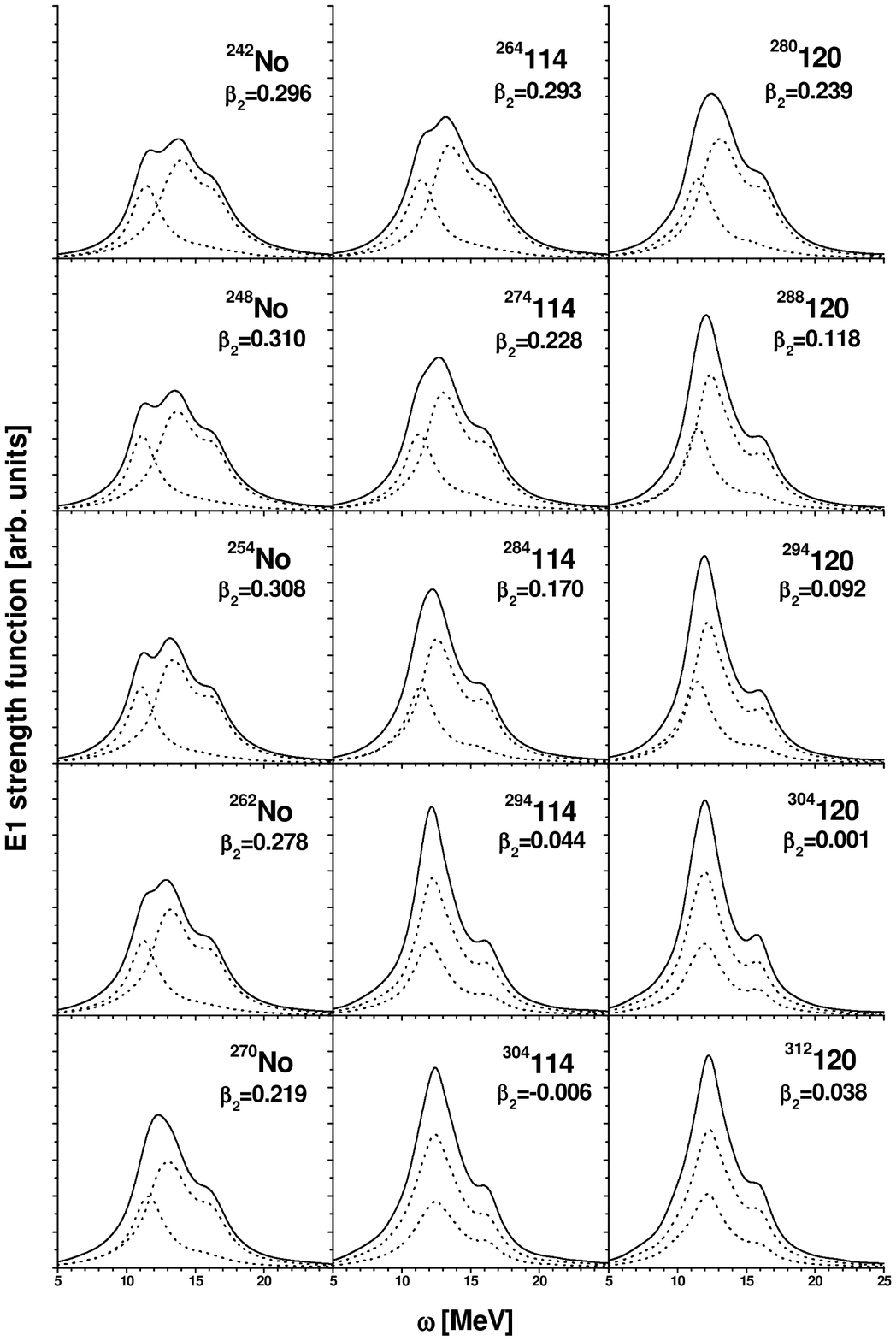}
\caption{\label{fig:super_E1}
Same as in Fig. 1 for isotopes of super-heavy nuclei No and
$Z$=114 and 120. The dimensionless proton quadrupole
deformation $\beta_2$ is indicated in the plots.
}
\end{figure}

\begin{figure}
\includegraphics[height=6.5cm,width=6.5cm,angle=-90]{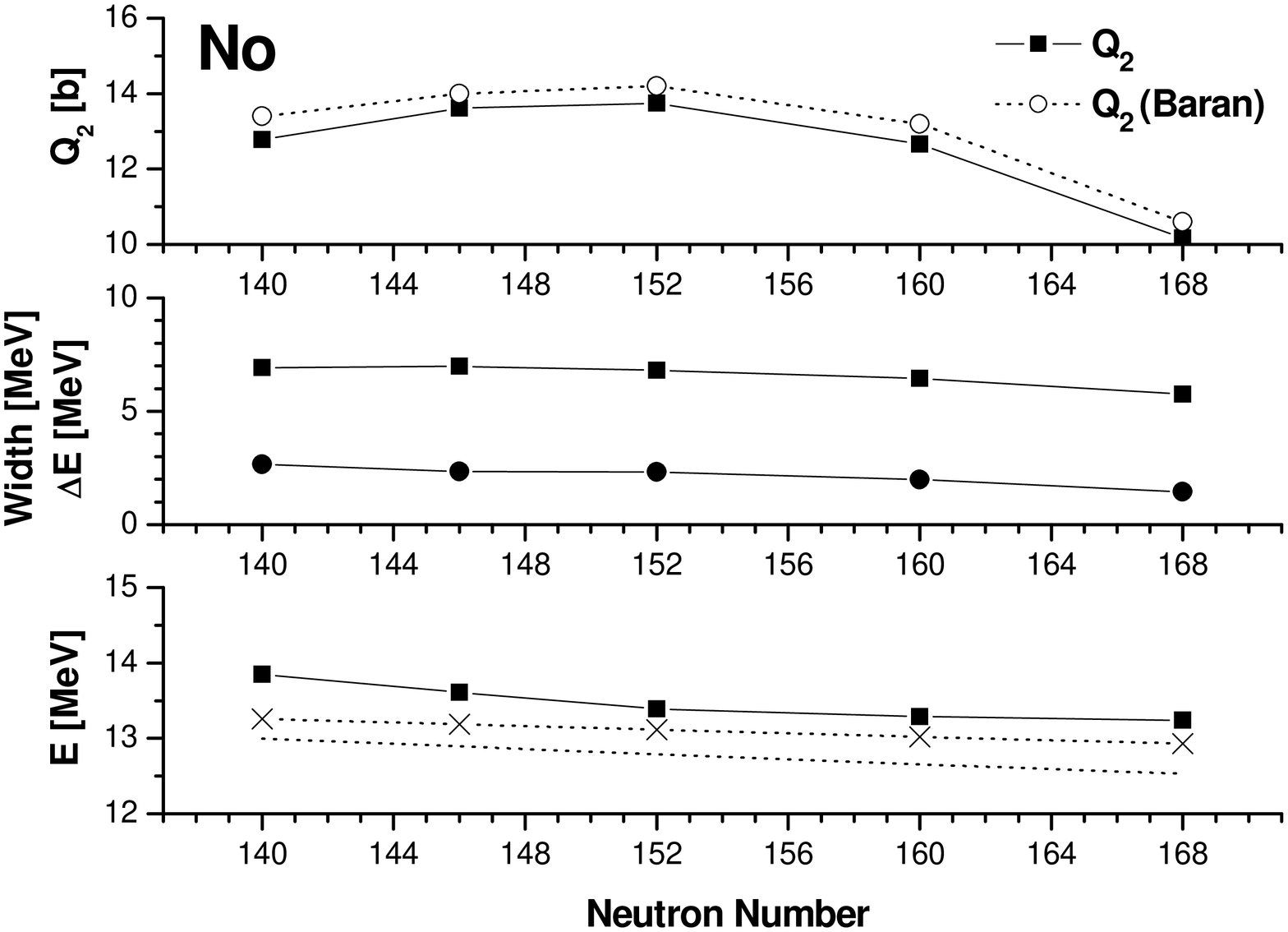}
\vspace{0.3cm}
\caption{\label{fig:no_trend}
Calculated characteristics of No isotopes as a function of
their neutron number. Upper panel: Quadrupole
moments $Q_2$ (black squares) as compared with the values
\protect\cite{baran} (open circles).
Middle panel: Widths $\Gamma$ (black squares) and
deformation splittings $\Delta E$ (black circles).
Lower panel: Energy centroids (black
squares) as compared with the estimates $E_{SJ}$
\protect(\ref{eq:SJ}) (dotted curve) and
$E_{BF}$ \protect(\ref{eq:BF}) (dotted curve with crosses).
}
\end{figure}

\begin{figure}
\includegraphics[height=6.5cm,width=6.5cm,angle=-90]{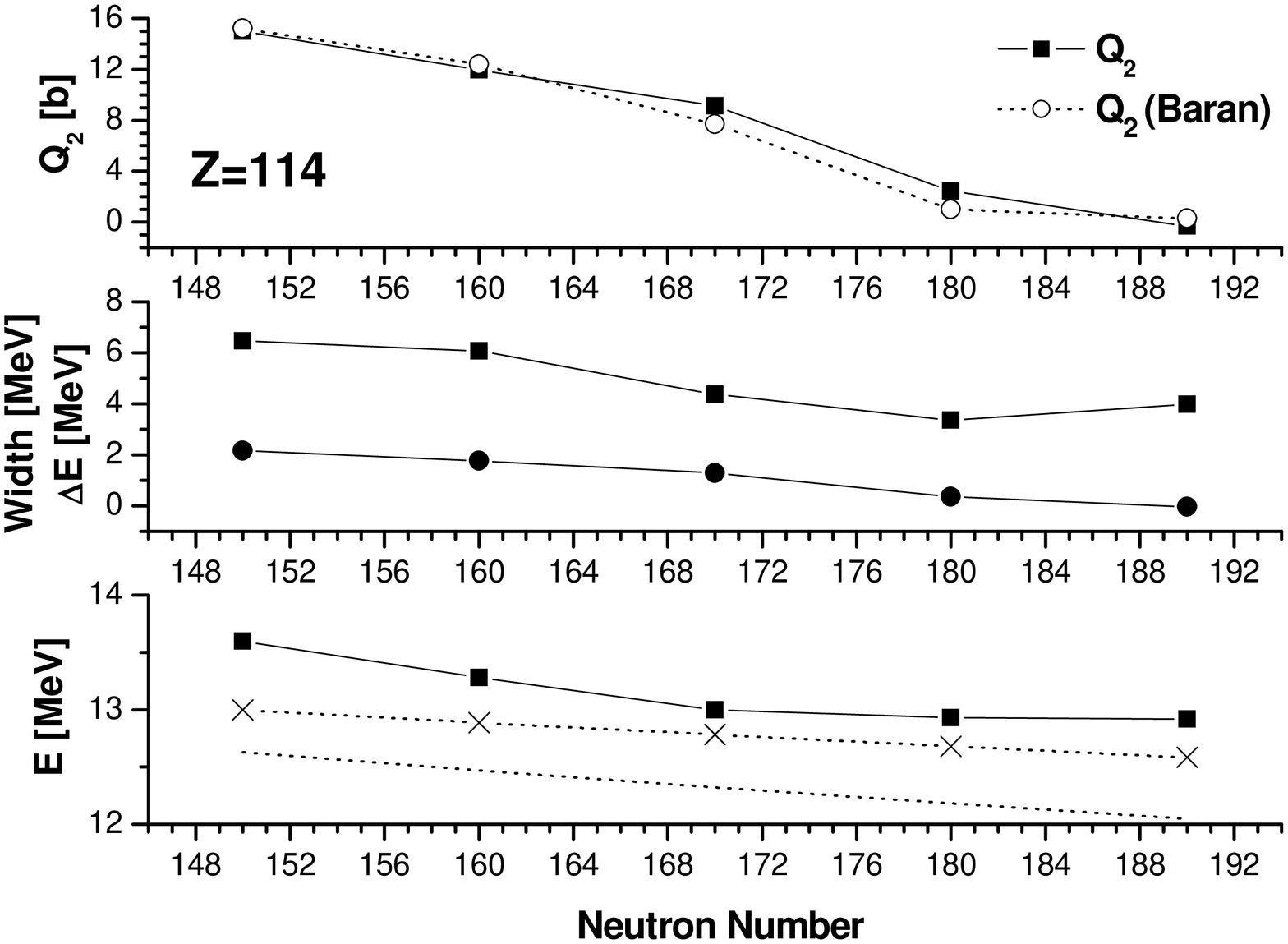}
\caption{\label{fig:114_trend}
Same as in Fig. 8 for $Z$=114 isotopes.
}
\end{figure}

\begin{figure}
\includegraphics[height=7cm,width=7cm,angle=-90]{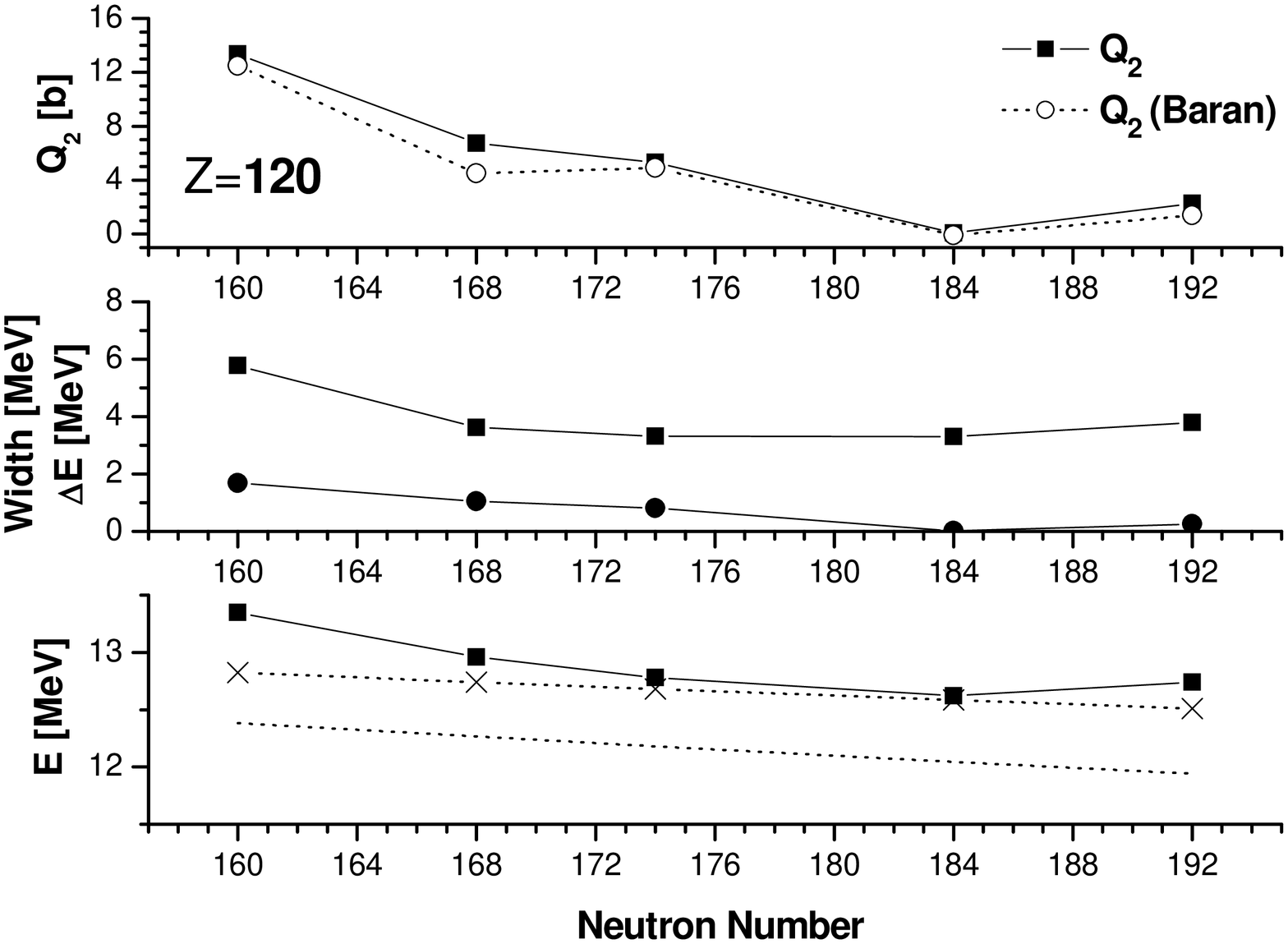}
\caption{\label{fig:120_trend}
Same as in Fig. 8 for $Z$=120 isotopes.
}
\end{figure}

The agreement of SRPA results with the experimental data in rare-earth and
actinide nuclei encourages its further application to super-heavy
nuclei.  SRPA results for super-heavy nuclei are exhibited in Figs.
\ref{fig:super_E1}-\ref{fig:120_trend}.

% Figure 7
%%%%%%%%%%%%

Fig. \ref{fig:super_E1} indicates that the GDR in this region
is generally similar to that in rare-earth and actinide nuclei.  In
particular, the resonance width correlates with the quadrupole
parameter $\beta_2$.  Furthermore, the middle panels in
Figs. \ref{fig:no_trend}-\ref{fig:120_trend} show that the direct
contribution of the deformation splitting $\Delta E$ to the resonance
width $\Gamma$ does not exceed $40\%$, as in rare-earth and
actinide nuclei.  Note that our quadrupole moments $Q_2$ from the
self-consistent calculations agree nicely with the values obtained
within the macroscopic-microscopic model \cite{baran}, see the upper
panels in Figs.  \ref{fig:no_trend}-\ref{fig:120_trend}.  The
agreement persist even in mass regions with large variations of
deformation.

% Figure 8
%%%%%%%%%%%%
% Figure 9
%%%%%%%%%%%%
% Figure 10
%%%%%%%%%%%%

At the same time, the GDR in super-heavy nuclei shows some
peculiarities. First, unlike the rare-earth nuclei, its energy is much
closer to the BF estimate (\ref{eq:BF}) than to (\ref{eq:SJ}), which
supports once again the treatment of the GDR in heavy nuclei as a
mixture of SJ and GT modes. Maximal deviations from both estimates
emerge at the ends of the isotopic chains, which is natural since
these estimates do not parameterize any isospin dependence.  Second,
as is seen from Figs. \ref{fig:no_trend}-\ref{fig:120_trend},
the decrease of the resonance energies with neutron number $N$ levels
off and is even reversed to some increase at the end of every isotope
chain. This can be explained by increase of the symmetry energy $E
\propto E_{sym}=a_{sym}(N-Z)^2/A^2$ at these neutron-rich edges. Note
that such a turnover is absent for lighter nuclei, e.g. in the chain
of Nd isotopes (A=134-158) explored earlier in
\cite{nest_kazim}. Probably this is because Nd isotopes do
not reach so large neutron excess as the super-heavy elements.

% Figure 11
%%%%%%%%%%%%
\begin{figure}
\includegraphics[height=9cm,width=7cm,angle=-90]{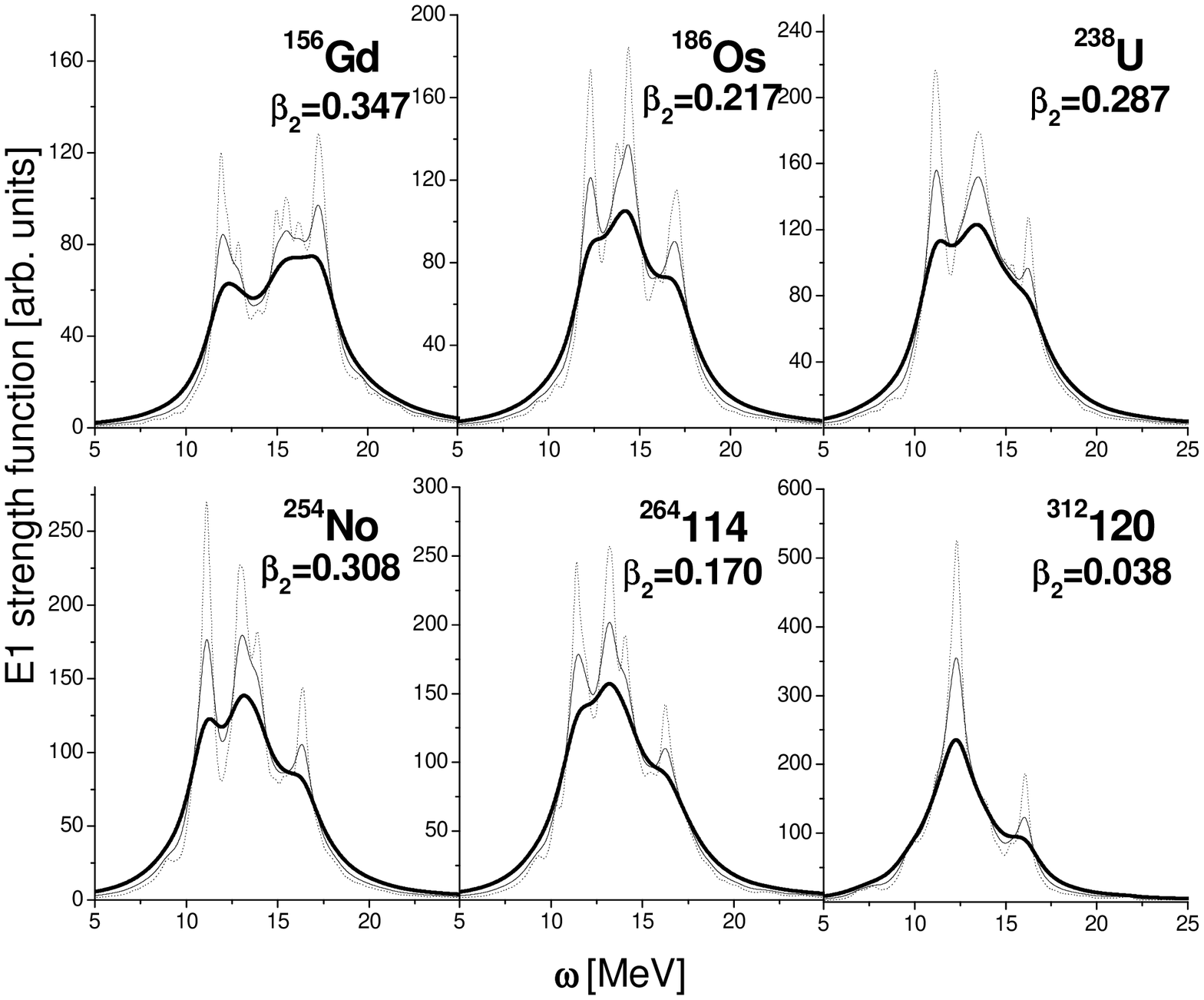}
\caption{\label{fig:var_aver}
The isovector dipole strength
calculated with different averaging in the strength function
(\protect\ref{eq:strength_function}):
$\Delta$ =0.5 MeV (dash curve), 1 MeV (thin solid curve),
and 2 MeV (bold solid curve). The parameters of the proton quadrupole
deformation are given for every nucleus.
}
\end{figure}

\subsection{GDR width and structure: general discussion}
\label{sec:width}

In Figure \ref{fig:var_aver}, the isovector dipole strengths
calculated with different Lorentz averaging parameter $\Delta$ are
compared for a representative set of nuclei.  It is seen that smaller
averaging, $\Delta$ =0.5 and 1 MeV, yields more fine structures, mainly
at the resonance peak, and leaves the total width almost unchanged. As
a rule, $\delta \Gamma=\Gamma(\Delta=2 \; \text{MeV})-\Gamma(\Delta=0.5
\; \text{MeV}) \le $ 1 MeV. Since $\Gamma (\Delta=$2 MeV) well reproduces the
experimental widths, one may associate the difference $\delta \Gamma$
to the smoothing effects omitted in the present RPA calculations
(coupling with complex configurations, escape widths). In fact, the
averaging with $\Delta=$2 MeV effectively mimics these effects. Since
$\delta \Gamma \ll \Gamma (\Delta=2 MeV)$, the deformation splitting
and Landau fragmentation (distribution of the collective strength
between nearby two-quasiparticle states) obviously dominate the total
width.  We estimate their contribution by 70-90$\%$, depending on the
nucleus and its shape.

Figure \ref{fig:var_aver} also shows that the averaging $\Delta=$2
MeV chosen in our calculations is indeed most suitable for the
comparison with the GDR experimental data (at least for the heavy nuclei
considered here). This averaging does not cause significant
artificial increase of the resonance width and, at the same time,
allows to suppress the structure  details which, in any case, are not
resolved in the experimental strength distributions.

%%%%%%%%%%%%%%%%%%%%%%%%%
%Table 2.
\begin{table}
\caption{\label{tab:Q2}
Calculated RPA widths $\Gamma_0$ and $\Gamma_1$ of the resonance branches
$\mu=0$ and 1, the sum $\Gamma_0 + \Gamma_1$,  and the
total width $\Gamma$. For every nucleus the deformation
parameters $\beta_2$ are given. The averaging is $\Delta = 2 MeV$.
For more details see the text.
}
\begin{tabular}{|c|c|c|c|c|c|}
\hline
Nucleus & $\beta_2$  & \multicolumn{4}{|c|}{ Widths [MeV]} \\
\cline{3-6}
&& $\Gamma_0$ &  $\Gamma_1$ & $\Gamma_0 + \Gamma_1$ & $\Gamma$ \\
\hline
 $^{156}$Gd   &0.347 & 3,05 & 4,74 & 7,79 & 7,69 \\
 $^{172}$Yb   &0.347 & 2,54 & 5,08 & 7,62 & 7,28 \\
 $^{186}$Os   &0.222 & 2,65 & 5,09 & 7,74 & 6,39 \\
 $^{238}$U    &0.287 & 2,62 & 5,11 & 7,73 & 6,86 \\
 $^{254}$No   &0.308 & 2,44 & 5,13 & 7,57 & 6,81 \\
 $^{264}$114   &0.293 & 2,74 & 5,11 & 7,85 & 6,47 \\
 $^{304}$114  &-0.006& 3,98 & 3,99 & 7,97 & 3,98 \\
 $^{304}$120  &0.001 & 3.32 & 3.26 & 6.58 & 3.30 \\
 $^{312}$120  &0.038 & 3,61 & 3,91 & 7,52 & 3,80 \\
\hline
\end{tabular}
%\end{ruledtabular}
%\end{center}
\end{table}
% Figure 12
%%%%%%%%%%%%
\begin{figure}
\includegraphics[height=5.0cm,width=6.5cm]{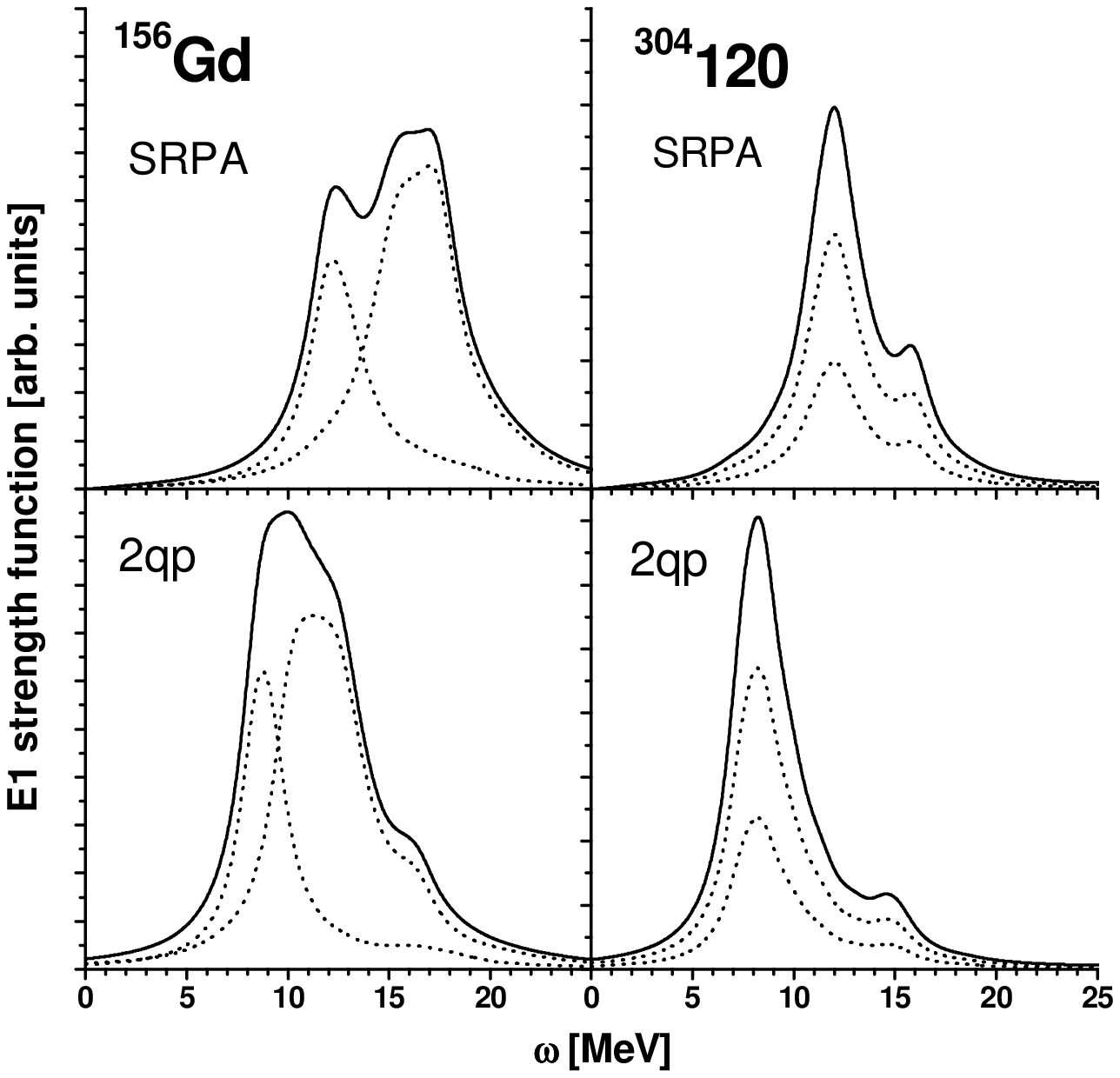}
\vspace{0.3cm}
\caption{\label{fig:branches}
The isovector dipole strength
calculated with smoothing $\Delta$ =2 MeV in deformed $^{156}$Gd
and spherical $^{304}$120. The upper panels show full SRPA results
(with residual interaction) and  the lower panels pure
two-quasiparticle (2qp) strengths (without residual interaction).
The separate branches $\mu$=0 (small) and $\mu$=1 (large)
are plotted by the dotted line and their sum by the full line.
}
\end{figure}

Besides the splitting of the GDR into two branches, the deformation
also results in a considerable redistribution of the strength within
every branch. In other words, the deformation severely influences the
Landau fragmentation itself. This effect is illustrated in Table II
and Fig. \ref{fig:branches}. Table II provides the widths of the
resonance branches $\mu=0$ and $\mu=1$ for a selection of nuclei. There
is a large difference between spherical and deformed nuclei.  In
spherical nuclei ($^{304}$114, $^{304}$120, $^{312}$120), we have
$\Gamma_0 \approx \Gamma_1 \approx \Gamma$, while in deformed nuclei,
$\Gamma_1 > \Gamma_0$ and their sum $\Gamma_0 + \Gamma_1$ roughly
covers the total width $\Gamma$. So, we see a strong deformation effect
within the branches themselves.  In order to analyze it, we
plot in the lower panels of Fig. \ref{fig:branches} the unperturbed
two-quasiparticle (2qp) strengths.  In deformed
nuclei ($^{156}$Gd) the inequality $\Gamma_1 > \Gamma_0$ appears
already in  2qp strengths, which shows that this is simply an effect of
the density of states. The residual interaction (SRPA case) does not cause
essential changes in the relation between $\mu=0$ and $\mu=1$ branches and
preserves $\Gamma_1 > \Gamma_0$ in deformed nuclei.
Besides showing the influence of deformation, Fig. \ref{fig:branches}
also illustrates the collective shifts from the unperturbed strengths
to the final ones.  The size of the shifts is, of course, related to
the actual force SLy6.  The shifts are 4.0  and 3.4 MeV
in $^{156}$Gd and $^{304}$120, respectively.

In most of the nuclei the calculations indicate a small shoulder at
the right flank of the resonance.  The heavier the nucleus, the
stronger the shoulder. At a small averaging width, the shoulder
becomes more pronounced and may even show up as a separate peak, see
e.g. results for $\Delta$=0.5 MeV in Fig. \ref{fig:var_aver}. This
effect is often absent, at least much less pronounced, in the
experimental data. The shoulder takes place in both deformed
and spherical nuclei and so is independent of deformation. It persists
not only for SLy6 but also for most of other Skyrme forces
\cite{nest_06,nest_07}.  As was shown in \cite{nest_07}, the effect is
caused by specific 2qp structures with high angular momentum (thus
large statistical weight) lying at the right GDR flank.
It is to be noted that RPA neglects some broadening
mechanisms (escape widths, coupling with complex, 2p-2h, etc,
configurations) which could soften these structures.
The shoulder can be further enhanced if the Skyrme force overestimates
the dipole collective shift \cite{nest_06,nest_07}.
The case calls for further detailed exploration.

\section{Conclusions}
\label{sec:summary}

The isovector giant dipole resonance (GDR) is systematically
investigated in rare-earth, actinide and super-heavy regions. The
study covers 37 nuclei altogether. Mainly axially deformed nuclei are
considered.  In all the nuclei, the calculated quadrupole moments
correctly reproduce the experimental data (rare-earth and actinide
regions) \cite{Ram_exp_Q2} or macroscopic-microscopic estimates
(super-heavy region) \cite{baran}.  The calculations are performed in
the framework of the self-consistent separable RPA approach (SRPA)
based on the Skyrme functional. The force SLy6 is used.

A satisfying agreement of the SRPA results with the available GDR
experimental data is found for 22 rare-earth and actinide
nuclei. Resonance energies and widths are well described.  The trends
of the peak energies are compared with simple estimates from
collective models. It is confirmed that the Steinwedel-Jensen (SJ) model
performs well for medium heavy nuclei while a mix of SJ and the
Goldhaber-Teller scenarios  is more appropriate for heavy nuclei.
Encouraged by these results, the survey is extended to super-heavy
nuclei where GDR in isotope chains with $Z$=102, 114 and 120 are
explored. The GDR in super-heavy nuclei is found to behave similar to
that in rare-earth and actinide nuclei.  The peak energies are,
again, better described by the mixed collective model, continuing the
trend from heavy nuclei.  A new feature in the super-heavy region is
that the peak energies turn from decrease to increase towards the
neutron rich ends of the isotopic chains (close to the drip lines).

We also analyze the GDR widths $\Gamma$. They are shown to be strictly
dominated (at least 70-90$\%$) by Landau fragmentation and deformation
contributions. The direct deformation
contribution through the splitting of the GDR into $\mu=0$ and
1 branches achieves 40\%. The Landau fragmentation is
severely affected by the deformation as well, which modifies the branch
widths and lead to $\Gamma_1 > \Gamma_0$.
The final step to agreement with experimental pattern is achieved by
Lorentz averaging the SRPA results, thus simulating missing broadening
mechanisms (e.g., escape widths and coupling with complex
configurations). A modest additional broadening of $\sim$ 1 MeV
suffices to reach realistic pattern.

\section*{Acknowledgments}
The work was partly supported  by DFG grant RE 322/11-1 as well as
the grants Heisenberg - Landau (Germany - BLTP JINR) and Votruba - Blokhintsev
(Czech Republic - BLTP JINR) for 2007 and 2008 years.
W.K. and P.-G.R. are grateful for the BMBF support under contracts
06 DD 139D and 06 ER 808. This work is also a part of the research
plan MSM 0021620859 supported by the Ministry of Education of the
Czech Republic. It was partly funded by Czech grant agency
(grant No. 202/06/0363) and grant agency of
Charles University in Prague (grant No. 222/2006/B-FYZ/MFF).

\end{document}